\documentclass[preprint]{vgtc}               %

\ifpdf%
  \pdfoutput=1\relax                   %
  \usepackage{graphicx}                %
  \DeclareGraphicsExtensions{.pdf,.png,.jpg,.jpeg} %
\else%
  \usepackage{graphicx}                %
  \DeclareGraphicsExtensions{.eps}     %
\fi%

\graphicspath{{figures/}{pictures/}{images/}{./}} %

\usepackage{microtype}                 %
\PassOptionsToPackage{warn}{textcomp}  %
\usepackage{textcomp}                  %
\usepackage{mathptmx}                  %
\usepackage{times}                     %
\usepackage{cite}                      %
\usepackage{tabu}                      %
\usepackage{booktabs}                  %

\onlineid{0}

\vgtccategory{Research}

\vgtcinsertpkg

\usepackage{enumitem}
\newcommand{\ie}{\textit{i.e.}}
\newcommand{\etal}{\textit{et al.}}
\newcommand{\eg}{\textit{e.g.}}
\newcommand{\para}[1]{\vspace{1mm}\noindent\textbf{#1}}

\newcommand{\stats}[2]{$m=#1,sd=#2$}
\newcommand{\fried}[2]{$\chi^2=#1, p=#2$}
\newcommand{\pValue}[1]{$p=#1$}
\newcommand{\userquote}[1]{\textit{``#1''}}

\newcommand{\re}[2]{\textcolor{black}{#1}}

\title{Towards an Understanding of \re{Distributed}{} Asymmetric \\ Collaborative Visualization on Problem-solving}

\author{Wai Tong\thanks{e-mail: wtong@connect.ust.hk}\\ %
        \scriptsize HKUST %
\and Meng Xia\thanks{e-mail: mengxia@andrew.cmu.edu}\\ %
     \scriptsize Carnegie Mellon University %
\and Kam Kwai Wong\thanks{e-mail: kkwongar@connect.ust.hk}\\ %
    \scriptsize HKUST %
\and Doug A. Bowman\thanks{e-mail: bowman@vt.edu}\\ %
    \scriptsize Virginia Tech %
\and Ting-Chuen Pong\thanks{e-mail: tcpong@ust.hk}\\ %
    \scriptsize HKUST %
\and Huamin Qu\thanks{e-mail: huamin@ust.hk}\\ %
    \scriptsize HKUST %
\and Yalong Yang\thanks{e-mail: yalongyang@vt.edu}\\ %
    \scriptsize Virginia Tech %
}

\teaser{
  \centering
  \includegraphics[width=0.8\linewidth]{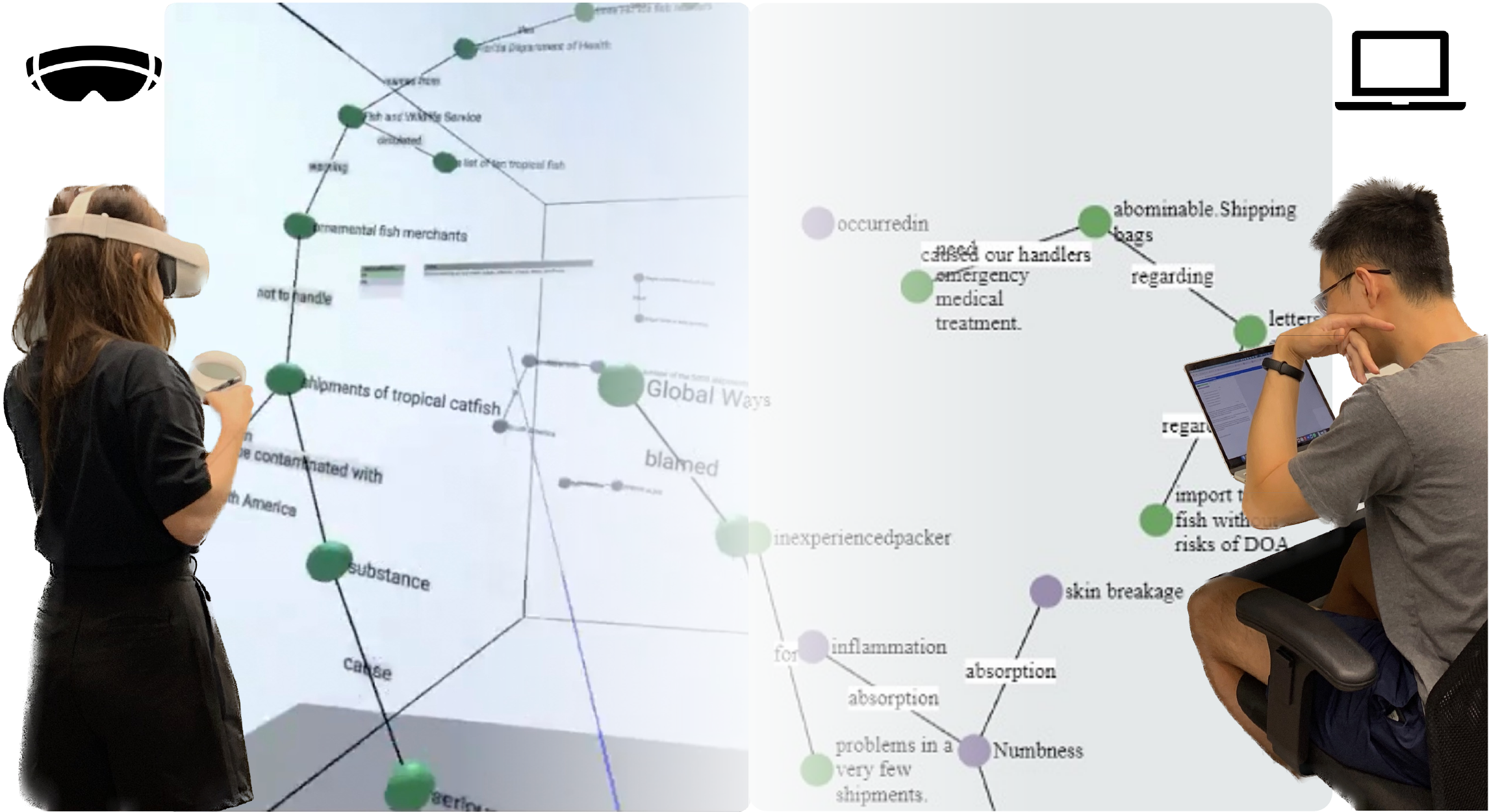}
  \caption{We studied the trade-offs of collaborative visualization for problem-solving in an asymmetric environment. This figure shows how two collaborators perceive and interact with visualizations using two different devices: VR (left) and PC (right).
  Visualizations are in different dimensions to adapt to different devices (i.e., 3D in VR and 2D on PC) and can be blended together (as envisaged in the center) with tailored techniques to support collaboration awareness.
  \vspace{1em}
  }
  \label{fig:teaser}
}

\abstract{
This paper provided empirical knowledge of the user experience for using collaborative visualization in a \re{distributed}{} asymmetrical setting through controlled user studies.
With the ability to access various computing devices, such as Virtual Reality (VR) head-mounted displays, scenarios emerge when collaborators have to or prefer to use different computing environments \re{in different places}{}. 
However, we still lack an understanding of using VR in an asymmetric setting for collaborative visualization.
To get an initial understanding and better inform the designs for asymmetric systems, we first conducted a formative study with 12 pairs of participants.
All participants collaborated in asymmetric (PC-VR) and symmetric settings (PC-PC and VR-VR).
We then improved our asymmetric design based on the key findings and observations from the first study.
Another ten pairs of participants collaborated with enhanced PC-VR and PC-PC conditions in a follow-up study.
We found that a well-designed asymmetric collaboration system could be as effective as a symmetric system. 
Surprisingly, participants using PC perceived less mental demand and effort in the asymmetric setting (PC-VR) compared to the symmetric setting (PC-PC).
We provided fine-grained discussions about the trade-offs between different collaboration settings.

} %

\CCScatlist{
  \CCScatTwelve{Human-centered computing}{Visu\-al\-iza\-tion}{Visu\-al\-iza\-tion techniques}{};
  \CCScatTwelve{Human-centered computing}{Collaborative and social computing}{Empirical studies in collaborative and social computing}{}
}

\begin{document}

\maketitle

\section{Introduction}
\re{Collaborative visualization~\cite{isenberg2011collaborative} becomes crucial in modern data workflows by providing a scalable solution to divide and conquer increasingly intensive data problems.
Moreover, many complicated data problems are intrinsically interdisciplinary, requiring people with different expertise, possibly from different locations, to work synchronously for analysis.
In particular, there is a growing interest in asymmetric collaboration (\ie, different collaborators use different computation devices~\cite{grandi2019characterizing}) for rightful reasons.
First, different computation devices have their affordability that can complement others. For example, Personal Computers (PCs) with mice and keyboards are the most familiar settings to control visual elements precisely. Virtual Reality (VR) Head-Mounted Displays (HMDs) have large display space~\cite{satriadi2020maps,lisle2020evaluating,hayatpur2020datahop}, embodied interaction~\cite{yang2020tilt,yang2020embodied,cordeil2017imaxes,tong2022exploring,ye2020shuttlespace}, and 3D rendering~\cite{bach2017hologram,yang2018maps,kraus2019impact,kwon2016study,wagner2019evaluating,chu2022tivee}.
Second, people possess different devices other than PCs. They have their preferences, accessibility, and spatial ability in using different devices~\cite{ouverson2021composite}.}{}

However, we lack an empirical understanding of people's asymmetric collaboration experiences of using visualization for problem-solving \re{remotely}{}.
Existing work has almost thoroughly investigated symmetric collaboration (\ie, collaborating on the same platforms) on different \re{devices}{platforms}, such as PCs~\cite{neogy2020representing,mahyar2014supporting,balakrishnan2008visualizations}, 
tabletops~\cite{tobiasz2009lark},
and VR~\cite{donalek2014immersive, lee2020shared, cordeil2016immersive,cavallo2019immersive}.
Most close to our work, Reski~\etal~\cite{reski2022empirical} tested the usability of asymmetric collaboration between PC and VR for \re{analyzing spatial data with map-based visualizations}{analytical tasks with spatial-data visualization using a map}. Results showed that participants generally considered asymmetric collaboration \re{usable}{acceptable}. Nonetheless, they did not compare the task performance and user experience between asymmetric and symmetric collaboration.
To this end, we asked
\textit{\re{how distributed asymmetric environments affect the task performance and user experience of collaborative visualization with node-link diagrams compared to symmetry ones}{how asymmetry affects the user experience of collaborative visualization compared to symmetry}.}

To answer the research question, we conducted a within-subjects study to compare the collaboration between an asymmetric (PC-VR) and two symmetric settings (PC-PC and VR-VR) with 12 pairs of participants.
\re{We chose the context of PC and VR since the PC is the most familiar computing device to most people, while VR has demonstrated promising results in immersive analytics~\cite{ens2021grand}.}{We chose the context of PC and VR since the PC is the most familiar computing device to most people, and high-quality VR has become increasingly popular and affordable. Particularly immersive analytics as an emerging research field demonstrates promising results of using VR for data analytics~\cite{ens2021grand}.}
\re{We designed the collaborative problem-solving task based on previous literature~\cite{balakrishnan2008visualizations,mahyar2014supporting}. The participants should build a node-link diagram from text documents and use it to answer analytical questions.}{We asked participants to complete a problem-solving task that was derived from the \re{literature}{litarture}~\cite{balakrishnan2008visualizations,mahyar2014supporting}, where they needed to build a \re{node-link diagram}{network visualization (or node-link diagram)} from the information they collected from text documents and use it to answer a set of analytical questions.}
\re{While designing the PC user interfaces for such a task has been widely explored~\cite{mahyar2014supporting, balakrishnan2008visualizations}, the VR counterpart for asymmetric collaboration has limited design guidance.}{Our PC user interface was primarily inspired by previous studies~\cite{mahyar2014supporting, balakrishnan2008visualizations}. Meanwhile, there is limited guidance in designing such a VR user interface for asymmetric collaboration.}
As the first step in exploring this design space, \re{we considered designing the VR interface metaphors aligned with the PC as closely as possible to reduce the communication burden between the PC and VR interfaces. Since we designed the VR interface by mirroring the PC interface, we called it \textit{mirrored VR}}{we designed the VR user interface as similar as possible to the PC (\ie, a mirrored design) so that the collaborators are more likely on the same page}.
We found that the asymmetric condition was as good as the symmetric conditions in performance and communication. However, participants generally did not prefer the mirrored design, \re{suggesting room for improvement to adapt the PC user interface to VR}{indicating that the user interface on PC was unsuitable for VR}. 

\re{The first formative study helped us}{The first study also serves as a formative study for us to} distill the necessary design requirements for collaborative visualizations and interactions in an asymmetric setting.
Based on the requirements, we significantly improved our asymmetric condition in the second study.
In particular, we enhanced the documents' presentation and VR interactions and provided additional awareness support for collaboration.
As a result, the interface in VR was no longer similar to the PC. Instead, the VR interface fully leveraged its intrinsic benefits, \ie, large display space, spatial memory and navigation, and intuitive embodied interaction.
We then conducted another within-subjects study to compare PC-PC and PC-VR with ten other pairs of participants.
As expected, most participants found no difficulty interacting with the updated VR user interface and performed better in reading documents.
Moreover, participants who used PC reported lower mental loads and efforts for completing the tasks in the asymmetric setting than in the symmetric setting.
We also observed that asymmetric settings might lead to implicit role assignments in collaborative work.

Overall, our contributions are three-fold: first, we conducted a formative study to extract design requirements of collaborative visualization in a \re{distributed}{} asymmetric setting; second, we developed a cross-virtuality web application to support both symmetric and asymmetric collaborative visualizations with PC and VR; third, we conducted a controlled user study investigating the trade-offs between asymmetric and symmetric collaborative visualizations\re{}{and offered a nuanced understanding of their user experience}. 

\section{Related Work}

\para{Data Visualization Beyond the Desktop.}
Data visualization has been used in traditional desktop environments for decades. 
With the development of different computation devices, researchers are moving beyond PC to various types of devices~\cite{roberts2014visualization}, for example, 
tabletops~\cite{isenberg2013data}, mobile devices~\cite{lee2018data}, and immersive HMDs~\cite{serrano_immersive_2022}.
Each device provides its advantage in data visualization. For example, mobile devices and smartphones are easy to carry out, and large displays and tabletops could easily support multiple people to analyze data on a shared surface.
As a result, visualization designers and researchers are passionate about investigating the usage of different devices for data visualization~\cite{munzner2014visualization}.
Recently, immersive HMDs have been rapidly developing, and data visualization researchers have started investigating the uniqueness of immersive displays and the potential benefits of data visualization. Three of the key benefits of immersive displays are large display space~\cite{satriadi2020maps,lisle2020evaluating,hayatpur2020datahop}, embodied interaction~\cite{yang2020tilt,yang2020embodied,cordeil2017imaxes,tong2022exploring,ye2020shuttlespace}, and 3D rendering~\cite{bach2017hologram,yang2018maps,kraus2019impact,kwon2016study,wagner2019evaluating,chu2022tivee}.

While there are different types of devices for data visualization, \re{they should work in complement rather than incompatible.}{these devices, especially immersive displays, are not going to replace but complement each other.} \re{For example, Wang~\etal~\cite{wang2020towards} proposed a hybrid PC and AR setup instead of PC or AR only for 3D data analysis. Hubenschmud~\etal~\cite{hubenschmid2022relive} also built a visual analytics system to analyze mixed reality user studies with an in-situ VR view and an ex-situ PC view.}{} Moreover, the performance in AR for data exploration varied on individual differences, such as the ability of spatial understanding and hand-eye coordination~\cite{bach2017hologram}. We envision that users will choose their preferred devices for data exploration and analytics in the future, determined by working environments and personal preferences.

\para{Collaborative Visualization.}
Data visualization has been widely used for collaboration in complex problem-solving and sensemaking tasks~\cite{balakrishnan2008visualizations,mahyar2014supporting}, environmental science~\cite{brewer2000collaborative}, and emergency management collaboration~\cite{wu2013supporting}. 
Isenberg~\etal~\cite{isenberg2011collaborative} defined collaborative visualization as ``\textit{the shared use of computer-supported, (interactive,) visual representations of data by more than one person with the common goal of contribution to joint information processing activities.}''
It improves the efficiency and accuracy of synchronous remote collaboration~\cite{balakrishnan2008visualizations}.

Previous studies investigated collaborative visualization in both PC-PC and VR-VR settings. For example, in the PC-PC setting, Mahyar and Tory~\cite{mahyar2014supporting} introduced externalization support by linking notes to the node-link diagram in data visualization to enhance communication and coordination.
In the VR-VR setting,
several studies investigated and evaluated collaborative visualizations from feasibility and performance perspectives~\cite{cordeil2016immersive,lee2020shared}.
Cordeil~\etal~\cite{cordeil2016immersive} conducted a user study to show that groups could do collaborative data analysis using 3D \re{graph visualizations}{graphs} in the immersive environment using both CAVE and VR. 
Lee~\etal~\cite{lee2020shared} continued to study collaboration using VR HMDs and showed that people could create and use 2D and 3D visualizations for data exploration in a collaborative virtual environment.
In addition to PC-PC and VR-VR, researchers also utilized Augmented Reality (AR) to support effective distributed collaboration with geographic data~\cite{mahmood2019improving} and co-located collaboration with multivariate data~\cite{butscher2018clusters}.

Most \re{collaborative visualization}{} research targeted symmetric collaboration, where all users work together in PC, VR, or AR only. 
However, each device type has its features and advantages, and people can access different devices.
Therefore, people could collaborate with different types of devices~\cite{ens2021grand}.
Yet, the asymmetry perspective remains underexplored in collaborative visualization.

\para{Asymmetric Collaboration.}
Asymmetric collaboration, interacting and viewing the content from different devices for collaboration~\cite{grandi2019characterizing}, has been studied in past years besides symmetric collaboration~\cite{sereno2020collaborative}. Asymmetry is considered not a challenge or limitation to overcome but rather a common social interaction~\cite{ouverson2021composite}.
\re{Different research has been conducted to enable collaboration with asymmetric devices.}{On the one hand, users could play different roles in asymmetric collaboration.} 
For example, ShareVR~\cite{gugenheimer2017sharevr} and ShARe~\cite{jansen2020share} were introduced to support communication between immersive users (using VR and AR headset, respectively) and non-immersive users (using projection and mobile devices), as well as remote users and local users~\cite{oda2015virtual}.
Additionally, Sugiura~\cite{sugiura2018asymmetric} introduced Dollhouse VR to support asymmetric collaboration between a customer and an architect. In particular, the architect could modify the room using a tabletop while the customers could immerse in experiencing different room settings. Similarly, Welsford~\etal~\cite{welsford2021spectator} implemented a collaborative system to allow users to spectate and communicate with another immersed VR HMD user on semi-immersed large displays.
Furthermore,
Grandi~\etal~\cite{grandi2019characterizing} found that participants in the asymmetric setting (VR-AR) performed significantly better than in the AR symmetric setting and similar to the VR symmetric setting. 

However, limited studies~\cite{reski2022empirical} have investigated asymmetric collaboration in the context of data visualization.
As the first step, we conduct studies to deepen our understanding of asymmetric collaborative visualization\re{, particularly about the asymmetrical use of devices}{}. Our result could benefit and guide the future design of asymmetric collaborative visualization systems.
\begin{figure*}[t]
\centering
\includegraphics[width=\textwidth]{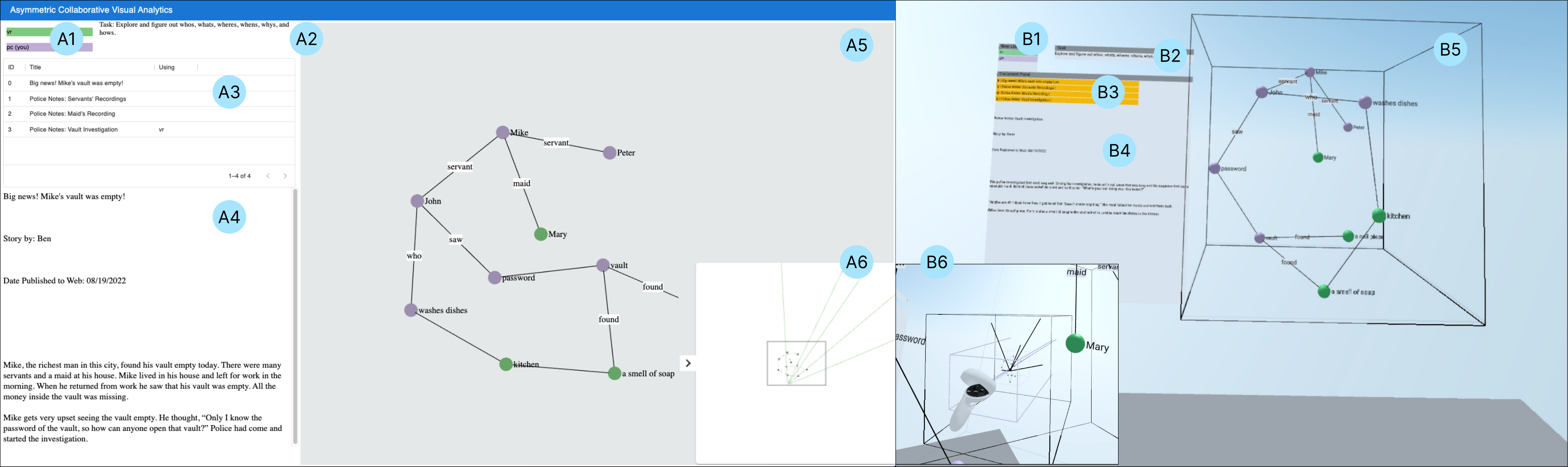}
\caption{The figure shows the prototype used in the formative study under PC (A1-A6) and VR (B1-B6). The application in both PC and VR consists of five components: the user list (A1, B1), the task description (A2, B2), the document list (A3, B3), the selected document (A4, B4), and the node-link graph (A5, B5). An overview window (i.e., a mini version of the node-link graph) is provided for both PC (A6) and VR (B6, attached to the left hand). A minimap is used on PC (A6), while a minicube is used in VR (B6). 
The line in black represents the current user's view frustum, while other colors represent the other users' view frustums with the corresponding color.}
\label{fig:interface}
\end{figure*}

\section{Formative Study}
\re{We conducted a formative study to understand how people collaborate using data visualization in asymmetric environments and 
the difference compared to collaborations in symmetric settings}{}.
We included two symmetric conditions (\ie, PC-PC and VR-VR) and one asymmetric condition (\ie, PC-VR). 
The study is approved by the Institutional Review Boards of the authors' university.

\subsection{\re{Participants}{UPDATE: Moved the participants from below to here}}
We recruited 24 participants (7 females and 17 males, aged 18-34), all undergraduate and graduate students, by sending recruitment emails and word-by-mouth to the local universities. 
18 out of 24 participants were computer science majors\re{. The other six were from biology, mechanical engineering, computer engineering, engineering, geographic information system, and architecture and design, respectively}{}.
For data visualization and VR experience, participants reported using interactive data visualizations (\eg, reading interactive infographic articles and using visualization dashboards) and VR HMDs (\eg, playing VR games and using VR applications) for more than two years~(12 for visualizations and 3 for VR), one to two years~(4 and 6), one month to one year~(4 and 4), less than one month~(1 and 9), and never~(3 and 2). 
The three participants with no visualization background were confirmed that they could read and create node-link diagrams.
The two participants without VR experience were confirmed to have experience in VR-related experience, \ie, using the PC version of VRChat and smartphone-based VR HMD.
Participants were formed into 12 pairs.
The familiarity between group members is as follows: very close to each other~(3), acquaintances~(5), and strangers~(4).
For collaboration experience, five participants mentioned that they have collaborative activities daily, 13 weekly, four monthly, and two less than monthly. 

\subsection{Task and Data}
Adapting from studies that investigate collaborative visualization~\cite{balakrishnan2008visualizations,mahyar2014supporting}, we used similar problem-solving tasks in our study.
For each condition, the group was required to read a set of documents and infer an illegal event from the documents.
Participants had to point out the details by answering whos, whats, wheres, whens, hows, and whys. They needed to create and manipulate a node-link diagram to facilitate solving the problem. 
We chose this task for the following reasons.
First, node-link diagrams do not require high visualization literacy~\cite{yang2023examples} and are commonly used in problem-solving and sensemaking tasks to reveal the relationship between different entities~\cite{lin2021taxthemis}.
Particularly for collaborations, node-link diagrams were helpful in externalizing users' thinking and improving the task performance and accuracy~\cite{mahyar2014supporting, balakrishnan2008visualizations}.
Moreover, 3D node-link diagrams were found effective in VR, providing better motivations for using VR in our study~\cite{ belcher2003using, ware1996evaluating, ware1994viewing, ware2005reevaluating}. With the motion and depth cues in VR, people can observe paths between nodes with lower error rates than PC~\cite{ware2005reevaluating}, 
potentially introducing benefits for asymmetric settings.

We used the Blue Iguanodon dataset~\cite{grinstein2007vast}, with approximately 1700 documents, from the VAST 2007 contest. 
This dataset was used in education at the graduate level, and thus the difficulty of the dataset is appropriate for undergraduates and graduate students~\cite{whiting2009vast}. 
The task was to find out unexpected illegal activities against wildlife from the provided documents.
To avoid the learning effect on the same dataset, we extracted three subplots with different illegal activities (\ie, drug trafficking, wildlife smuggling, and bioterrorism), where each subplot was mapped to one of the three conditions for a group.
Considering the time of the user study session~\cite{yoghourdjian2020scalability},
we retrieved and provided the most-related documents for each subplot. 
Finally, each subplot contains six documents with similar total word counts (\ie, 813, 779, and 805, respectively).

\subsection{Design and Implementation}
\label{ssec:networkvis}
To support the use of node-link diagrams with documents for the aforementioned tasks, we designed a prototype system with two main views: documents view and node-link graph view for both PC and VR, as shown in \autoref{fig:interface}.
Our PC user interface was primarily inspired by previous studies~\cite{mahyar2014supporting, balakrishnan2008visualizations}.
Meanwhile, since there was limited guidance in designing such a VR user interface, we introduced the \textit{mirrored VR} as the first step in exploring this design space and lowering the communication efforts between two different computation environments. Mirrored VR means the VR application was designed similar to the PC application in terms of interaction and visual interface. 
The major differences were the dimensionality of the working environment and the \re{graph visualization}{graph} (\ie, 2D on PC and 3D on VR).
This decision was made based on empirical evidence demonstrating the benefits of 3D \re{graph}{network} visualization over 2D \re{graph}{network} visualization in VR~\cite{kwon2016study,yang2018origin}.

\para{Documents View.} Documents view contains four components: the user list, the task description, the document list, and the selected document. The user list (\autoref{fig:interface}; A1, B1) shows all users inside the current room. Each user is assigned a unique color to indicate their created nodes and view frustums. 
The task description (\autoref{fig:interface}; A2, B2) shows the task the users should work with. This view is explicitly added 
to remind participants about the key elements, \ie, who, what, when, where, how, and why, to be answered. The document list (\autoref{fig:interface}; A3, B3) shows all the available documents, including the document ID, the document title,
and the current reading status of all users.
Users can select a document from the document list to be shown below (\autoref{fig:interface}; A4, B4).

\para{Node-link Graph View.} Node-link graph view (\autoref{fig:interface}; A5, B5) is a place for users to create and read the node-link diagram.
The 2D and 3D graph visualization are shown to PC and VR users, respectively.
The visualization is shared among all users because it improves performance, encourages collaborators to use the tool, and facilitates discussions~\cite{balakrishnan2008visualizations}.
The visualization contains:
\begin{itemize}[itemsep=0pt,topsep=0pt,partopsep=0pt, parsep=0pt]
    \item 
    \textit{Node:} Users can \textit{move}, \textit{add}, \textit{modify}, \textit{merge}, and \textit{delete} nodes in the \re{graph visualization}{graph}. 
    They can add nodes for entities found in the documents by selecting the texts in the document and placing them in the graph visualization.
    The nodes' color encodes the creating user.
    
    \item 
    \textit{Link:} Users can define the relationship between two nodes by adding the links. 
    Similarly, the links' labels are extracted from the documents. 
    They can \textit{add}, \textit{modify}, and \textit{delete} links in the \re{graph visualization}{graph}.
\end{itemize}
To maintain a similar graph layout during collaboration, we started from computing the 3D layout of the node-link diagrams.
We applied a 3D force-directed graph layout for VR users and then projected it to the 2D X-Y plane for PC users.
To avoid overlapping specific nodes, we applied the force-directed layout with the same parameter again to the 2D projected layout, where the positions of non-overlapping nodes are unchanged.
In addition, we provided an overview window for the node-link graph view. 
We used a minimap on PC (\autoref{fig:interface}(A6)) and World In Miniature~\cite{stoakley1995virtual} \re{}{(\ie, a mini cube)} (\autoref{fig:interface}(B6)) in VR.

\para{Interaction designs.}
Users can interact with the application using \textit{mouse} on PC and \textit{VR controller} in VR.
\begin{itemize}[itemsep=0pt,topsep=0pt,partopsep=0pt, parsep=0pt]
    \item 
    \textit{Mouse:} Users can \textit{pinch} and \textit{drag} to zoom and pan the \re{graph visualization}{graph} on a PC. Then, by \textit{right-clicking} the node, a context menu shows up and allows users to perform the deletion, merge, move, refer to documents, and highlight.
    \item 
    \textit{VR Controller:} Users can scale and rotate the \re{graph visualization}{graph} using \textit{two-hand manipulation}. In addition, users can select words by \textit{pressing} the trigger button and \textit{dragging} through the text on the document. Similar to PC, we provided the same context menu in VR. However, users need to long-trigger the node for 1 second to ``right-click'' the node.
\end{itemize}

\begin{figure}
\centering
\includegraphics[width=\columnwidth]{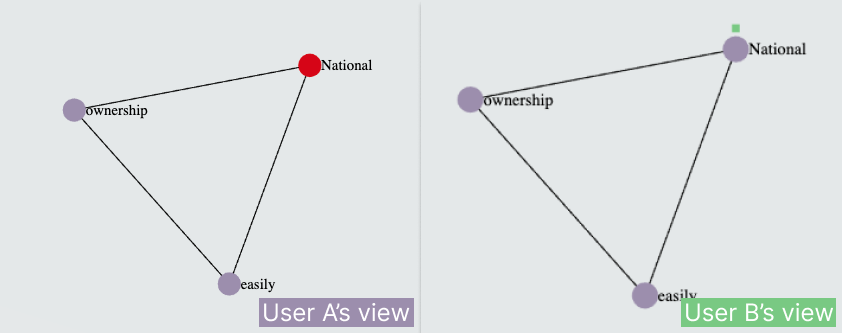}
\caption{The figure shows the highlight feature when selecting nodes. When user A selected a node (node in red color), user B could see a square (or a cube in VR) with user A's representative color appearing on top of the selected node.}
\label{fig:highlight}
\vspace{0.5em}
\end{figure}

\para{Awareness support for collaboration.} We supported awareness in the document list view, the node-link graph view, and the overview window.
First, we displayed all users' document reading status in the document list view by listing users' names in the ``Using'' column in \autoref{fig:interface}(A3).
Second, we added a visual cue to indicate all users' node selection status to support awareness in the \re{graph visualization}{graph} while minimizing the interference to the analysis~\cite{neogy2020representing}. As shown in \autoref{fig:highlight}, when a user selects a node, all other users see a square (or a cube in VR) with the user's representative color appearing on top of the selected node.
Third, we encoded the node using its creator's representative color, similar to prior works~\cite{mahyar2014supporting}.
Lastly, we provided head rays and view frustums in the overview window for awareness~\cite{yang2020embodied}. On PC, the head rays and the view frustums of the other collaborators are projected (\autoref{fig:interface}(A6)), while they are shown directly in VR~\cite{piumsomboon2019effects}.
Their colors represent the different users. 

\para{Implementation.} 
We implemented both PC and VR applications using React.js, d3.js, three.js, and WebXR.
To support communication between different applications, we adapted the client-server architecture. 
We implemented the server using Node.js and gRPC for fast data transfer. All operations and layout calculations were done on the server, and the result was sent back to the clients by streaming.
We used the force-directed layout algorithm on 2D and 3D node-link diagrams from an extended version of d3-force\footnote{\url{https://github.com/vasturiano/d3-force-3d}}.
\re{The code is open-sourced at \url{https://github.com/asymcollabvis/asymcollabvis}}{}.

\begin{figure}[t]
\centering
\includegraphics[width=.8\columnwidth]{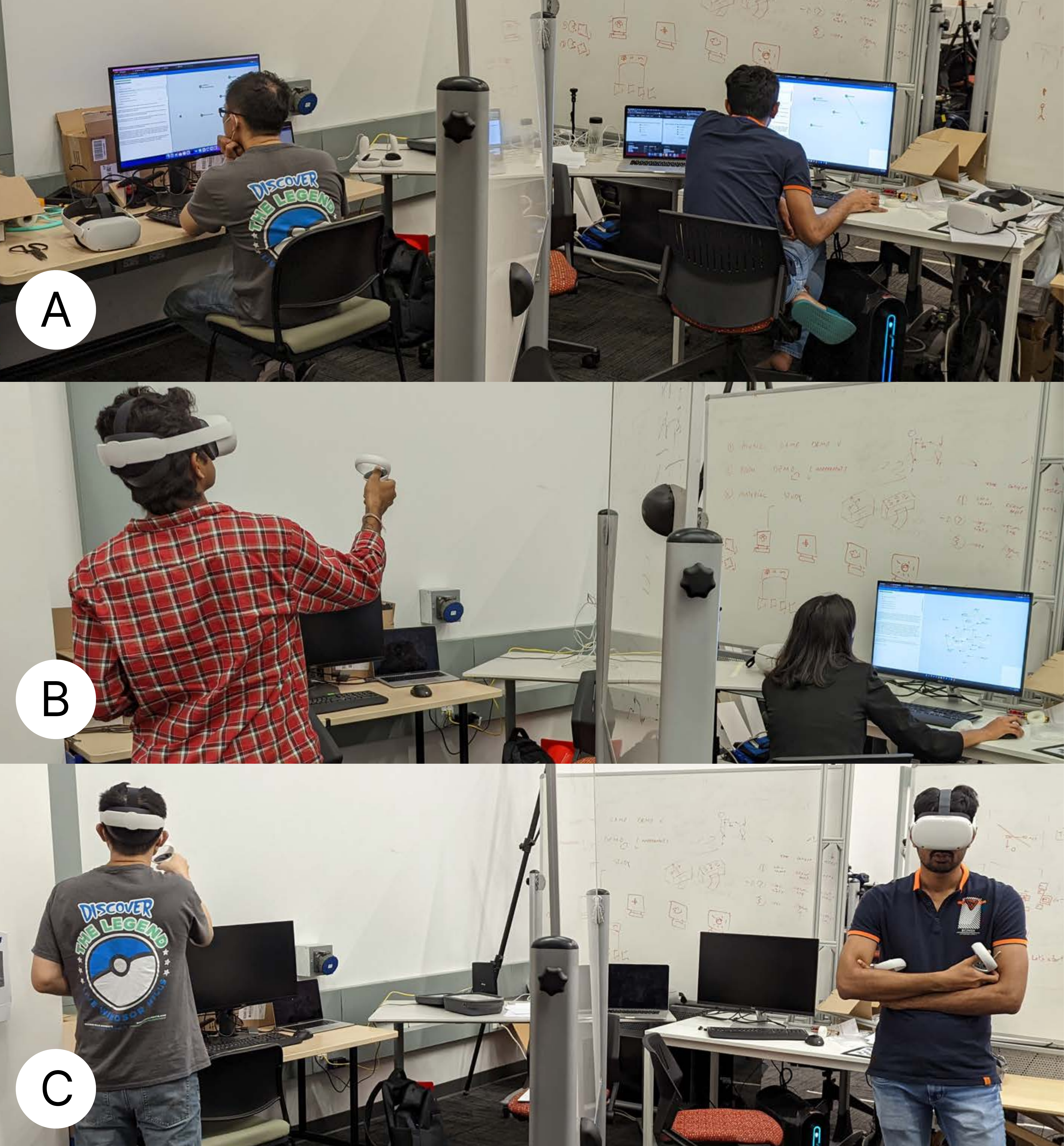}
\caption{The figure shows two participants working together in PC-PC (A), PC-VR (B), and VR-VR (C) conditions. A \re{standard office whiteboard}{} was placed in the middle to simulate a remote setting.
}
\vspace{0.5em}
\label{fig:setting}
\end{figure}

\subsection{Apparatus and Setting}

As shown in \autoref{fig:setting}, our experimental setup included three conditions: PC-PC (A), PC-VR (B), and VR-VR (C). 
We put a \re{standard office whiteboard}{board} between them to simulate a distributed collaboration scenario, but we allowed them to talk to each other similar to using online communication tools.

\subsection{Procedure}
The study consisted of four parts: introduction, training, main study, and debriefing. It lasted for about 120 minutes, and a \$20 USD Amazon eGift card was given to each participant as compensation. The study had been audio-recorded and logged for system operations.

\para{Introduction (5 minutes).} A pair of participants were welcomed and instructed about the purpose of the study, the duration of the study, as well as the setup. The participants then read the consent form and signed it if they agreed and wanted to proceed with the study. Participants were informed that they could take breaks at any time they wanted, and they could withdraw at any stage. 

\para{Training (30 minutes).} The conductor introduced the PC and VR visualization system according to established guidelines~\cite{yang2023examples}. Since some participants had little experience with VR HMDs, the PC system was first introduced before the VR system to reduce the learning curve. The PC and VR interactions were demonstrated using the same data and task to show what the participants could perform in both systems. Then, participants were asked to practice the introduced functions to complete a training task in both PC and VR. The training task, figuring out the underlining crime, was the same as the main task but with a more straightforward dataset with only four documents (318 words in total).

\para{Main Study (75 minutes).}
Each pair of participants tried all three conditions in the main study. We counterbalanced the sequence of conditions using the balanced Latin Square method~\cite{bradley1958complete} and controlled the sequence of the dataset.
A general background of the task was given to the group.
Before each condition started, we reminded the participants of the background and the task. We encouraged them to 1) collaborate as earlier as possible, 2) distribute the documents, and 3) discuss how to use the \re{graph visualization}{graph} together.
After each condition, participants had a 5 minutes break. Meanwhile, they needed to answer the NASA Task Load Index (TLX) questionnaire~\cite{sandra2006nasatlx} and the Behavioral Engagement questionnaire~\cite{biocca2001networked}, for task load and social presence measures, adapted from previous study~\cite{grandi2019characterizing}.
A short interview was conducted regarding the collaboration strategy, benefits, challenges, and suggestions for each condition. 

\para{Debriefing (10 minutes).} After completing the main study, we presented individuals with a questionnaire to rank the conditions. We then discussed their preferences and potential inspirations from participating in the study and using the prototype systems.

\begin{figure}
\centering
\includegraphics[width=\columnwidth]{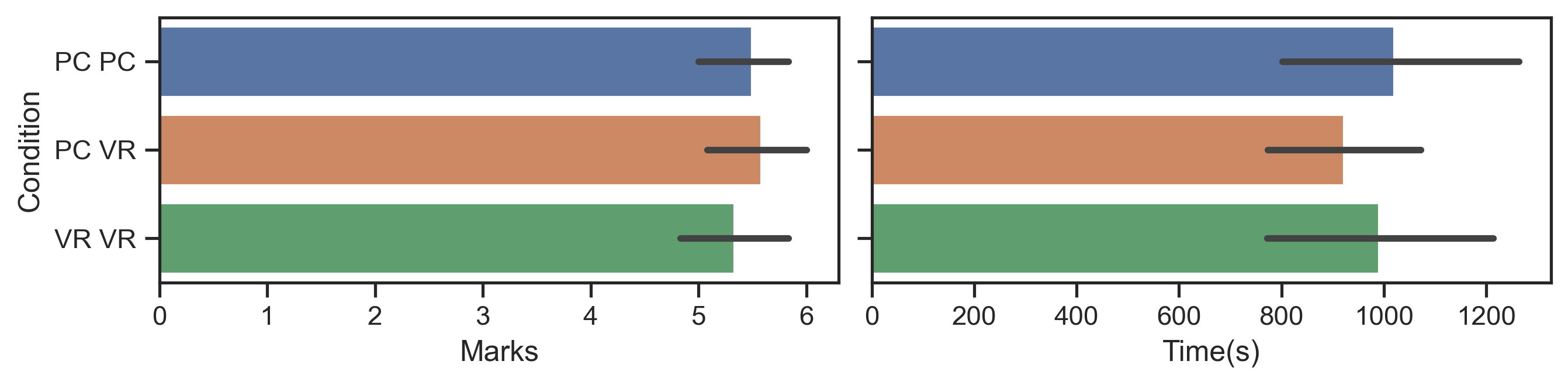}
\caption{The figure shows the mean and 95\% confidence intervals of average task completion score and time for three conditions.}
\label{fig:result_mark_time}
\end{figure}

\subsection{\re{Preference}{Qualitative Feedback}}
\re{17 participants preferred collaborating in symmetric environments (16 in PC-PC and 1 in VR-VR), and 7 participants preferred asymmetric collaboration more than symmetric settings. 
We further analyzed the reasons from their qualitative feedback using affinity diagramming~\cite{hartson2012ux}. We denoted P1-24 as the 24 participants and G1-12 as the 12 pairs.
}{We received feedback for three different conditions from short discussions after each condition and a semi-structured interview during the debriefing session.}

\para{Symmetric collaboration.} All participants described the symmetric collaboration as the \userquote{most familiar} collaboration environment, especially for PC-PC. For PC-PC, participants also liked the PC interface to be \userquote{easy to control}.
VR-VR were described as \userquote{immersive} and \userquote{cool}, yet \userquote{having a high learning curve} and \userquote{difficult to interact and read documents}.

\para{Asymmetric collaboration.} PC-VR was generally acceptable to participants, having 7 participants preferred asymmetric collaboration more than both symmetric settings. Five participants (P8, P15-17, P21) suggested that using different devices could leverage the benefit of both devices during the collaboration. For example, P21 commented \userquote{it is natural to interact in 2D and it's easier to visualize if both are in 3D but with PC-VR we get the advantages of both.}
Based on the asymmetric benefits provided, four participants (P4-6, P8) stated asymmetric collaboration encouraged them to communicate and divide the work.
For example, P6 explained \userquote{PC users can easily access information, while VR users help guided meetings and discussions.}
Lastly, three participants (P18, P20, P23) were positive that the asymmetric collaboration helped participants collaborate using their preferred devices. For example, P18 pointed out that \userquote{I think the benefit mainly is because he (P17) got comfortable with the PC. So he was putting nodes and then I got comfortable with the VR. So I was putting nodes and making connections between them. ... I saw him adding a lot of nodes. And that kind of motivated me to connect the nodes together.} It echos prior work that asymmetric setting improves motivation~\cite{drey2022towards}.
However, for collaboration experience, P1 and P11 reasoned that they were uncomfortable with asymmetric working space. P1 stated that she was not comfortable when collaborating in different environments. P1 added \userquote{it is uncomfortable because I cannot help my partner since he is in a different environment.} P14 thought that \userquote{it may be easier to collaborate with someone when we are both in the visual space in the same mode}.

\subsection{Key findings}
We have collected the task completion time and accuracy, and responses for task load and social engagement in each condition. 
For task completion accuracy, we collected the answer to \textit{who}, \textit{what}, \textit{where}, \textit{when}, \textit{why}, and \textit{how}~\cite{grinstein2007vast}. One mark was given to one correct answer, and the full mark was six.
Below we reported the key findings found in the formative study.

\begin{figure}
\centering
\includegraphics[width=\columnwidth]{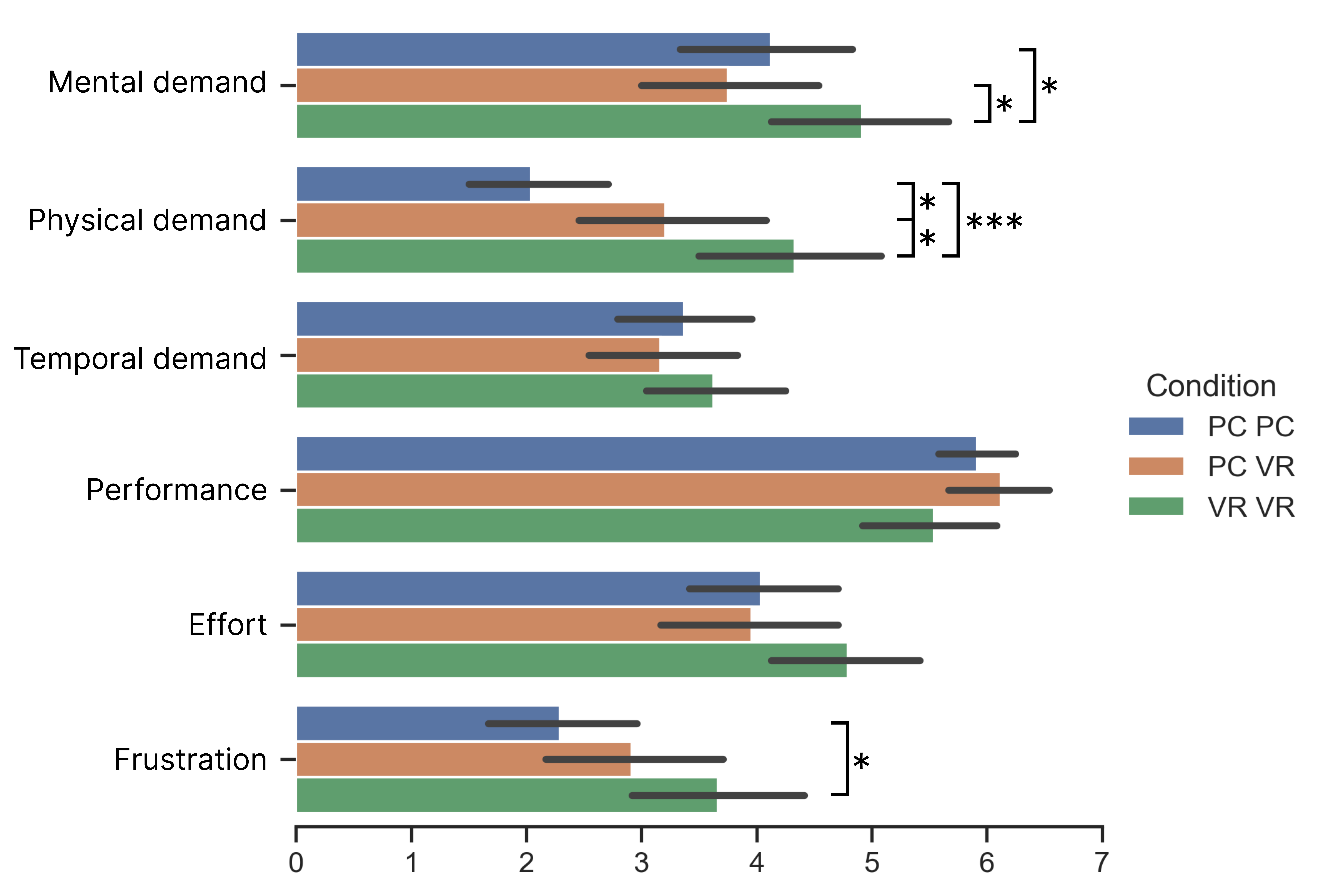}
\caption{The figure shows the mean and 95\% confidence intervals of the NASA task load questionnaire for three conditions.}
\label{fig:result_tl}
\end{figure}

\para{F1. Asymmetric settings did not affect performance and communication.} In the formative study, the users were provided with almost identical user interface and interaction no matter what platform they are using. The only asymmetry is the dimensionality of the \re{graph visualization}{graph}. 
By comparing the data collected using the Friedman test, we did not find significant differences in correctness (\fried{0.167}{0.920}) and time spent (\fried{0.667}{0.717}) between all conditions as shown in \autoref{fig:result_mark_time}. However, while we expected that the physical load would be significantly increased (\fried{22.6}{1.25e^{-5}}) when involving VR, we found a significant difference (\fried{8.22}{0.0164}) in the mental load between asymmetric and VR symmetric settings as shown in \autoref{fig:result_tl}. One possible reason is that VR users could be supported by PC users, who have a more familiar control.

\para{F2. Verbal communication and sharing data visualization are important for asymmetric settings.} Based on the interview feedback, participants reported that they did not find any difference when his/her partner was using different devices. One of the major reasons received is that participants were still able to talk to each other. For example, P12 said \userquote{I think there is no difference of what the others used for collaboration. One reason is that we can talk. It might be different when we are not able to talk to each other.} Moreover, participants mentioned that using shared data visualization was also one important thing for people to collaborate in different environments. For example, P22 said \userquote{for me since I could see what she drew. So I think, in that case, collaboration was pretty simple.}

\para{F3. Optimize both ends for collaboration.} By having the \re{mirrored}{low asymmetric} setting, we observed that participants have no difference in communication and using the \re{graph visualization}{graph} compared to symmetric settings. However, participants suggested features to improve their experience in using their devices. First of all, participants suggested customizing the input using keyboards on the PC (G10). Moreover, half of the groups (6/12) pointed out that the current design in VR did not use all the benefits of VR. For example, G2 and G7 mentioned that the 3D space was not used much in VR, and G9 commented that \userquote{it's really hard for us to have every point or maybe add links specifically}. As a result, the system should enhance the abilities of each platform, specifically the graph creation ability in PC and the utilization of the space and interaction in VR. 

\para{F4. Explicit visual cues for collaboration awareness are preferred.} 
Participants requested more explicit awareness cues. They thought the current way to direct the other collaborator's attention was discrete and not obvious. P14 said that \userquote{The node is highlighted when my partner selects it, but I couldn't see how my partner's attention moves continuously in the graph (visualization).} P21 suggested that \userquote{I think it will be better if we can see others' cursors on the screen (PC)}. Furthermore, both participants in G3 commented that the PC user completely ignored the VR user if the VR user did not say something. Participants also preferred to see others in the form of avatars in VR instead of viewing the frustum in the overview window for awareness.

\para{F5. Pitfalls.}
\re{Though the first study provided insightful findings, there were some pitfalls. First, 
we observed that the preference for VR-VR was unexpectedly low. The major reason was that it was hard to find participants with high expertise in VR for an in-person study, especially during the COVID-19 pandemic. Participants reported a high learning curve and unfamiliar controls as barriers for them to work together using the VR interface. 
Second, most participants tended to communicate verbally instead of using visualization for problem-solving. One reason was that the current task might be too easy due to the small data size; therefore, the users could memorise or quickly search for the entities across different documents. Participants reported only a medium effort according to the task load questionnaire (PC-PC: \stats{4.04}{1.68}; PC-VR: \stats{3.96}{1.90}; VR-VR: \stats{4.79}{1.64}). We might experience a ceiling effect in the study.}{}

\begin{figure}[t]
\centering
\includegraphics[width=\columnwidth]{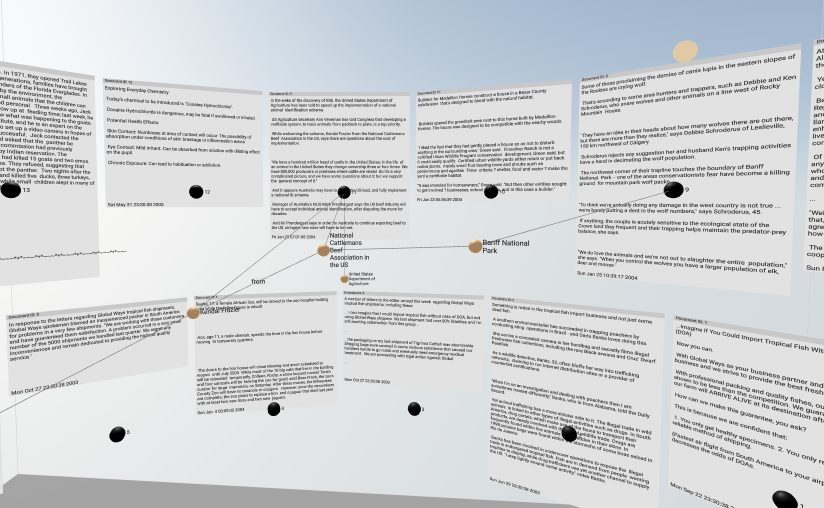}
\caption{\re{Documents are listed in a semicircle layout in VR. The nodes in black color represented each document and were created initially. A default link is added to the created node (the brown node) with the currently selected document to maintain a better spatial arrangement.}{Users can create nodes anywhere in the space and view the nodes in front of the related documents.}}
\label{fig:immersiveEnv}
\end{figure}

\section{Designing Asymmetric Collaborative Visualization for Problem-solving}
Based on the key findings from the first study, 
we sought to optimize the individual experience
to investigate whether this can positively affect the collaborative experience.

We improved the prototype and made the following changes to optimize information presentation and interaction on both ends (F3) and enhance collaboration awareness (F4).
Specifically, for F3, we have (1) added typing feature to provide custom labeling in PC condition; (2) arranged the documents distributed in the space to enable spatial sensemaking in VR; (3) added embodied interaction for adding/deleting nodes/links in VR. For F4, we made awareness cues for collaboration explicit by visualizing the cursor on both PC and VR and showing the view frustums to the VR space.

\subsection{Optimize on both PC and VR}

\para{Provide custom labeling in PC.}
In general, PC provided the most familiar input devices, mouse and keyboard, for precise control and so as in this study. For example, most of the participants stated that they could manipulate the \re{graph visualization}{graph} on PC better than in VR \re{because they were new and not familiar with using the VR controller to interact with graph visualization}{}.
Moreover, participants suggested custom label input would help use the \re{graph visualization}{graph} for problem-solving.
Therefore, we enabled the use of the keyboard for label creating and renaming on PC. In detail, we added an input textbox for users to enter the label on both nodes and links rather than just selecting the words from the documents.

\para{Enable spatial sensemaking in VR.}
Participants in the formative study suggested having a more immersive VR experience. Moreover, recent research has investigated the benefit of using the large display space in VR for sensemaking and data analysis~\cite{lisle2020evaluating,ens2022immersive}. For example, by utilizing spatial references, users increased their memory of the content of the documents~\cite{yang2020virtual}. Moreover, Hayatpur~\etal~\cite{hayatpur2020datahop} proposed DataHop to support spatial data exploration and showed that spatial mapping the workflow encouraged users to explore the data and provide a clear cluster of analysis without confusion. 
Therefore, we optimized spatial usage for problem-solving and sensemaking.

We have first arranged the documents distributed in the space in a semi-circular shape, as shown in \autoref{fig:immersiveEnv}. Semi-circular layouts are the most common layout in immersive visualization and document reading~\cite{satriadi2020maps,hayatpur2020datahop}. 
In addition, we combined the document space and the visualization space.
Users were able to create nodes anywhere in the space. 
To strengthen the spatial relationship between the document and the \re{graph visualization}{graph}, we provided one node
in front of each document  
and externalized the relationship between the created node and the document by adding a default link between the created node and the currently selected document nodes for a better spatial arrangement (\autoref{fig:immersiveEnv}). 

\begin{figure}[t]
\centering
\includegraphics[width=\columnwidth]{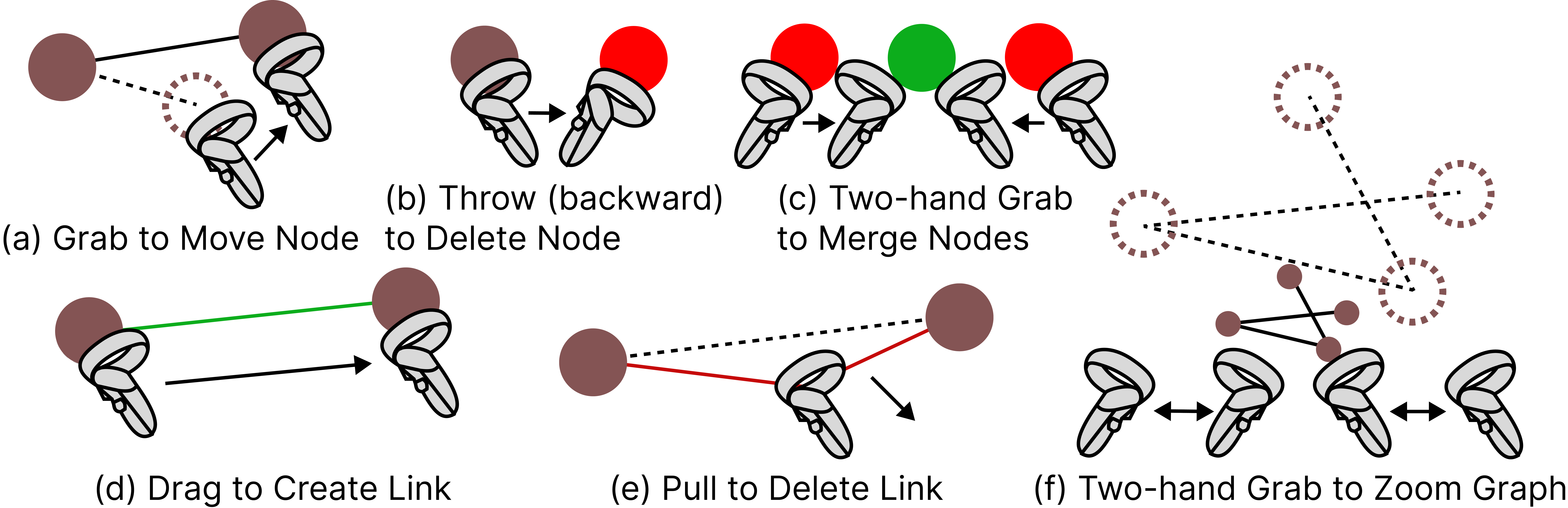}
\caption{Embodied interaction provided in the VR applications: users can (a) grab the node to move, (b) throw a node behind to remove a node,(c) grab two nodes using two controllers together to merge nodes, (d) drag a node to another node to create a link, (e) pull a link back to remove the link, and (f) grab the \re{graph visualization}{graph} using two controllers closer to zoom in. \re{The green color indicates nodes/links to be added. The red color indicates nodes/links to be deleted. The arrow indicates the movement of the controllers.}{}}
\vspace{0.5em}
\label{fig:interaction}
\end{figure}

\para{Support Embodied Interaction in VR.}
Besides utilizing the unlimited space, we deployed more natural and embodied interactions, instead of WIMP-like interaction, for node-link graphs in VR. 
Embodied interactions are promised to benefit immersive analytics~\cite{ens2022immersive}. Recent works~\cite{cordeil2017imaxes,yang2020tilt,ye2020shuttlespace,tong2022exploring,yang2020embodied} utilized embodied interactions to increase the effectiveness and engagement during data analysis.
Similarly, as illustrated in \autoref{fig:interaction}, we adapted metaphors of balls to the nodes and rubber bands to the links.
Users could move and grab the node around and delete the node by throwing the node, similar to moving and throwing a ball away.
For links, users can add a link between two nodes by first picking one end and pulling to the other end, just like using a rubber band to hold multiple objects together. To remove a link, users could grab it and pull it under the link break, similar to breaking a rubber band.

\begin{figure}
\centering
\includegraphics[width=\columnwidth]{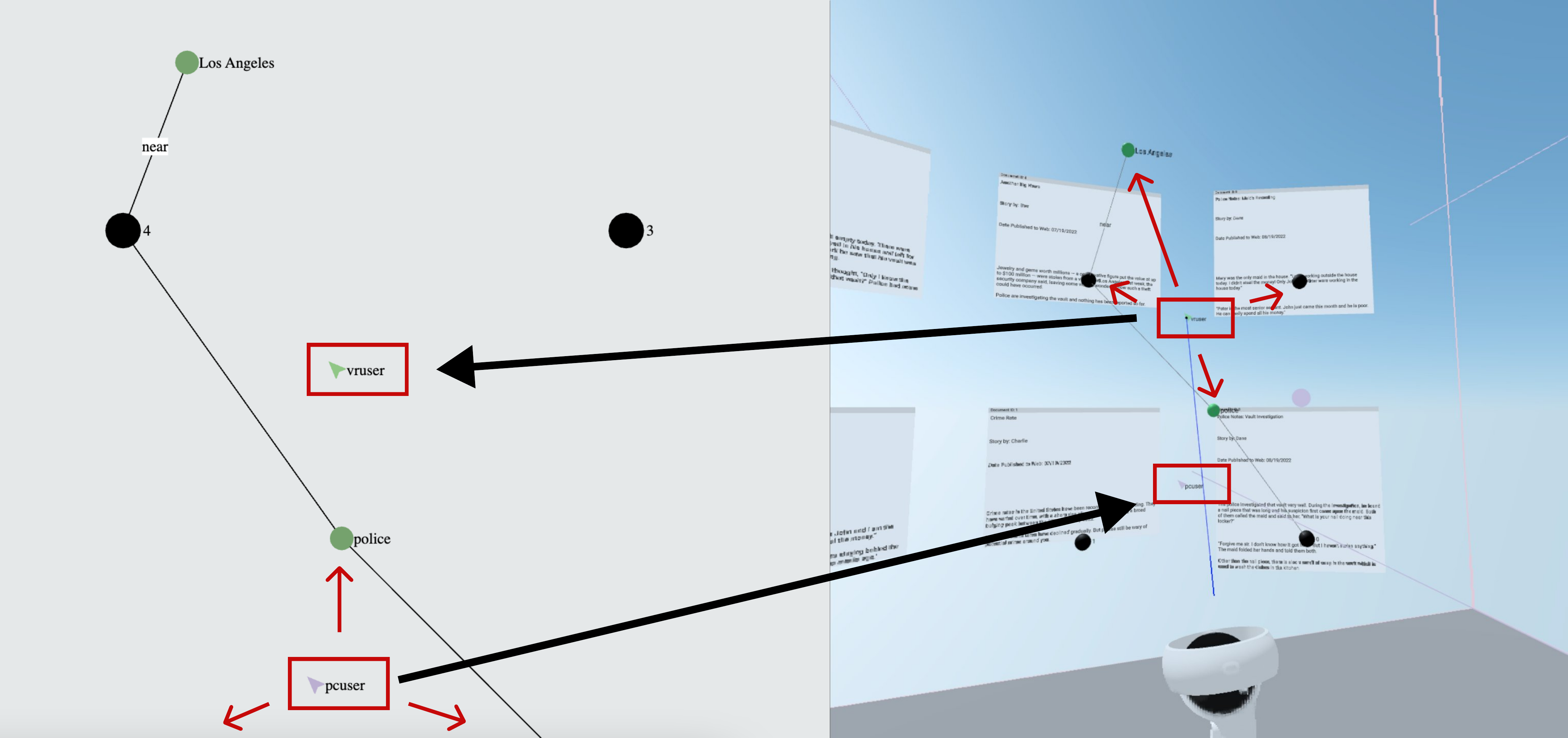}
\caption{Users can see the depth-adaptive cursor of other users in either PC or VR. The label of the cursor represents the user of the cursor. Red arrows indicate the natural neighbors found using the Voronoi diagram. We send the calculated node's weight and ID to the other platform (following the black arrows). The other platform can relocate the cursors by using the position of the received ID and the corresponding weight.}
\label{fig:depthcursor}
\end{figure}

\subsection{Make Awareness Explicit on PC and VR}
In the first study, participants did not use the highlight features to see the selected node of the collaborator. Instead, participants requested the visual cues of collaborators' attentions more directly in the \re{graph visualization}{graph} for awareness and communication.
Moreover, higher asymmetry in PC and VR applications provides more communication challenges since the interface is more different. 
We decided to implement a more explicit and always-on awareness cue, also commonly used in commercial collaboration software, \re{such as Miro and Microsoft Whiteboard}{}, a cursor, to both 2D and 3D environments, as shown in \autoref{fig:depthcursor}.

Adapted from the depth-adaptive cursor for using mouse input in VR~\cite{zhou2022depth}, we presented all users depth-adaptive cursors applicable to both platforms by interpolating the cursor position in both 2D and 3D based on the collaborators' cursor positions, the viewpoints, and the nodes. 
Specifically, we first found the natural neighbors of the cursor in PC or the ray in VR by computing the Voronoi diagram. After that, we calculated the weight of each neighbor using Laplace weights. Finally, the weight and neighbor were sent to all other users.
By interpolating the position of the nodes of the other users' cursor in PC or the ray in VR, we relocated the cursors in the other environment.

Moreover, we added the view frustum directly inside the space to show the view area of the other users in VR. Originally, we offered the view frustum in the overview window in VR (\autoref{fig:interface}(B6)). However, we observed that most users hid the mini cube in VR and reported that it was hard to understand. Participants mentioned that they wanted to see others in the VR space. Therefore, we moved the frustums and head rays from the overview window to the working space. The frustums and head rays can serve as a minimal avatar and improve users' awareness of collaborators' existence and behaviors.
\section{User Study}
\label{sec:study}
We conducted a controlled, within-subject study to explore and evaluate how refined asymmetric conditions affect collaboration in problem-solving tasks. We analyzed the pros and cons of the symmetric environment (\ie, PC-PC) and the refined asymmetric environment (\ie, PC-VR) for collaboration. 
\re{Based on the formative study, we found that most participants verbalized their findings rather than reading the graph visualization for collaboration because they could memorize the findings given a small number of documents.}{}
Therefore, we increased the task difficulty and maintained each condition session to last an hour on average.
We did not include the VR-VR condition because the different levels of experience in VR had introduced confounding factors to the previous study. 
Another reason for excluding VR-VR is to control the experiment time in a reasonable range.
Also, finding two participants with sufficient and similar VR experiences is challenging, especially during the \re{COVID-19}{} pandemic.
In the PC-VR condition, we ensured the VR participants had some VR experience.
Without VR-VR, participants did not feel uncomfortably fatigued during the study.

\subsection{Participants}
We recruited 20 participants (6 females and 14 males, aged 18-39), all undergraduate and graduate students, by sending recruitment emails and word-by-mouth. 
12 participants were computer science majors\re{, with the other ones from engineering~(6), computer engineering~(1), and biology~(1)}{}.
For data visualization and VR experience, participants \re{reported using}{stated they had used} interactive data visualizations \re{and VR HMDs}{} for more than two years~(7 for visualizations and 1 for VR), \re{}{seven used it for }one to two years~(7 and 5), \re{}{three used it for }one month to one year~(3 and 4), \re{}{one used \re{it}{} for }less than one month~(1 and 5), and \re{}{two }never~(2 and 5)\re{}{ used \re{it}{}}. 
The two participants \re{with}{had} no visualization background \re{were}{. They further} confirmed that they could read and create node-link diagrams.
The five participants \re{without VR}{who did not have any} experience\re{}{using VR} were not assigned\re{}{to} the VR headset \re{to avoid biased results due to their unfamiliarity with VR devices}{for potentially reducing performance and rating due to the unfamiliarity of VR devices}.
Participants were assigned into ten groups of two.
\re{The familiarity between group members is as follows:}{There are three groups that were} very close to each other~(3), \re{acquaintances~(3)}{three groups that knew each other but are not familiar with}, and \re{strangers~(4)}{four groups that \re{had}{were} never met before}.
For collaboration experience, three participants mentioned that they have collaborative activity daily, one weekly, three monthly, and three less than monthly. 

\subsection{Task and Data}
We chose the same task and two of the datasets (\ie, drug trafficking and wildlife smuggling) from the preliminary study. To increase the complexity of the task and the use of \re{graph visualization}{graph}, we appended extra background or unrelated documents to the original datasets. The total number of documents has changed from 6 to 15 per condition, with word counts of 2583 and 2518, respectively.

\subsection{Apparatus, Setting, and Procedure}
The apparatus and setting were the same as in the first study. 
However, each pair was required to complete tasks with only two conditions, and we provided more time for each condition (\ie, about 35 minutes) due to increased task complexity. \re{Same as in the first study, we counterbalanced the sequence of conditions using the balanced Latin Square method~\cite{bradley1958complete} and controlled the sequence of the dataset.}{}

\subsection{Measures and Research Questions}
To help us to answer the research question: \textit{how does asymmetry affect the user experience of distributed collaborative visualization}, we mainly consider performance and collaboration. Specifically, we focus on five aspects: task efficiency, perceived task load, \re{graph visualization}{graph} usage, behavior engagement, and communication process, adapted from previous studies~\cite{mahyar2014supporting}. Lastly, we also collected participants' preferences towards the two conditions by ranking.

To measure performance, we mainly focused on task efficiency, perceived task load, and \re{graph visualization}{graph} usage based on prior studies~\cite{balakrishnan2008visualizations, mahyar2014supporting}.
Same as in the formative study, for task efficiency, we collected the answer to \textit{who}, \textit{what}, \textit{where}, \textit{when}, \textit{why}, and \textit{how}. We also collected the user-perceived task load using the NASA TLX questionnaire~\cite{sandra2006nasatlx}.
Furthermore, different from the formative study, for \re{graph visualization}{graph} usage, we logged the interaction behavior for all sessions. All operations mentioned in \autoref{ssec:networkvis} were logged.

To measure collaboration, we looked at behavior engagement and the communication process. We measured the behavior engagement using the responses from Networked Minds Measure of Social Presence~\cite{biocca2001networked}.
To analyze the communication process, we audio-recorded the sessions and coded the transcripts into six categories: discussion of hypotheses (DH), referring to the visualization (RV), coordination (CO), seeking awareness (SA), verbalizing findings (VF), and asking questions about another group member's findings (QF)~\cite{mahyar2014supporting}.
\section{Result}
We ran the Wilcoxon signed rank test to test whether there is a significant difference between PC-PC and PC-VR. Moreover, we were also interested to see individual differences. \re{We further denoted the \textbf{individual} who used PC in PC-PC and PC-VR conditions as \textbf{I}1 and the individual who used different devices (\ie, PC in PC-PC and VR in PC-VR) as I2.}{} We ran another Wilcoxon signed rank test to check the pair-wise difference between the following pairs (\ie, \re{I1 in PC-PC (PC-PC$_{I1}$) and I1 in PC-VR (PC-VR$_{I1}$), P$_{I2}$ in PC-PC (PC-PC$_{I2}$) and P$_{I2}$ in PC-VR (PC-VR$_{I2}$), and P$_{I1}$ in PC-VR (PC-VR$_{I1}$) and P$_{I2}$ in PC-VR (PC-VR$_{I2}$)}{PC-PC, PC-VR, PC-PC, PC-VR}). To analyze qualitative data, we used affinity diagramming~\cite{hartson2012ux} for organizing the feedback.

\begin{figure}
\centering
\includegraphics[width=\columnwidth]{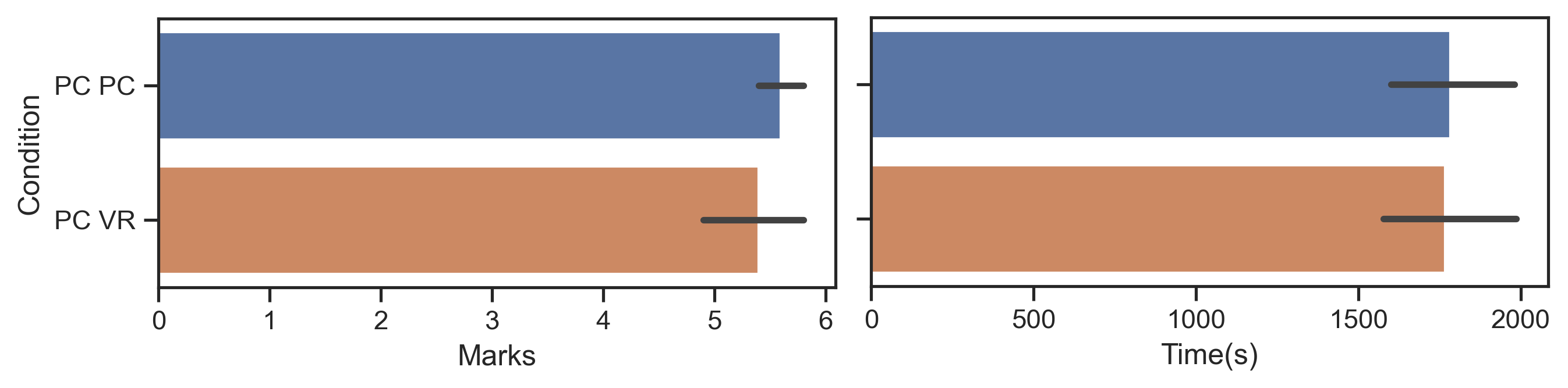}
\caption{The figure shows the mean and 95\% confidence intervals of average score and completion time for two conditions.}
\label{fig:result2_mark_time}
\end{figure}

\subsection{Performance}
In terms of accuracy and time, we did not find a significant difference between PC-PC and PC-VR (\autoref{fig:result2_mark_time}). 

\begin{figure}
\centering
\includegraphics[width=\columnwidth]{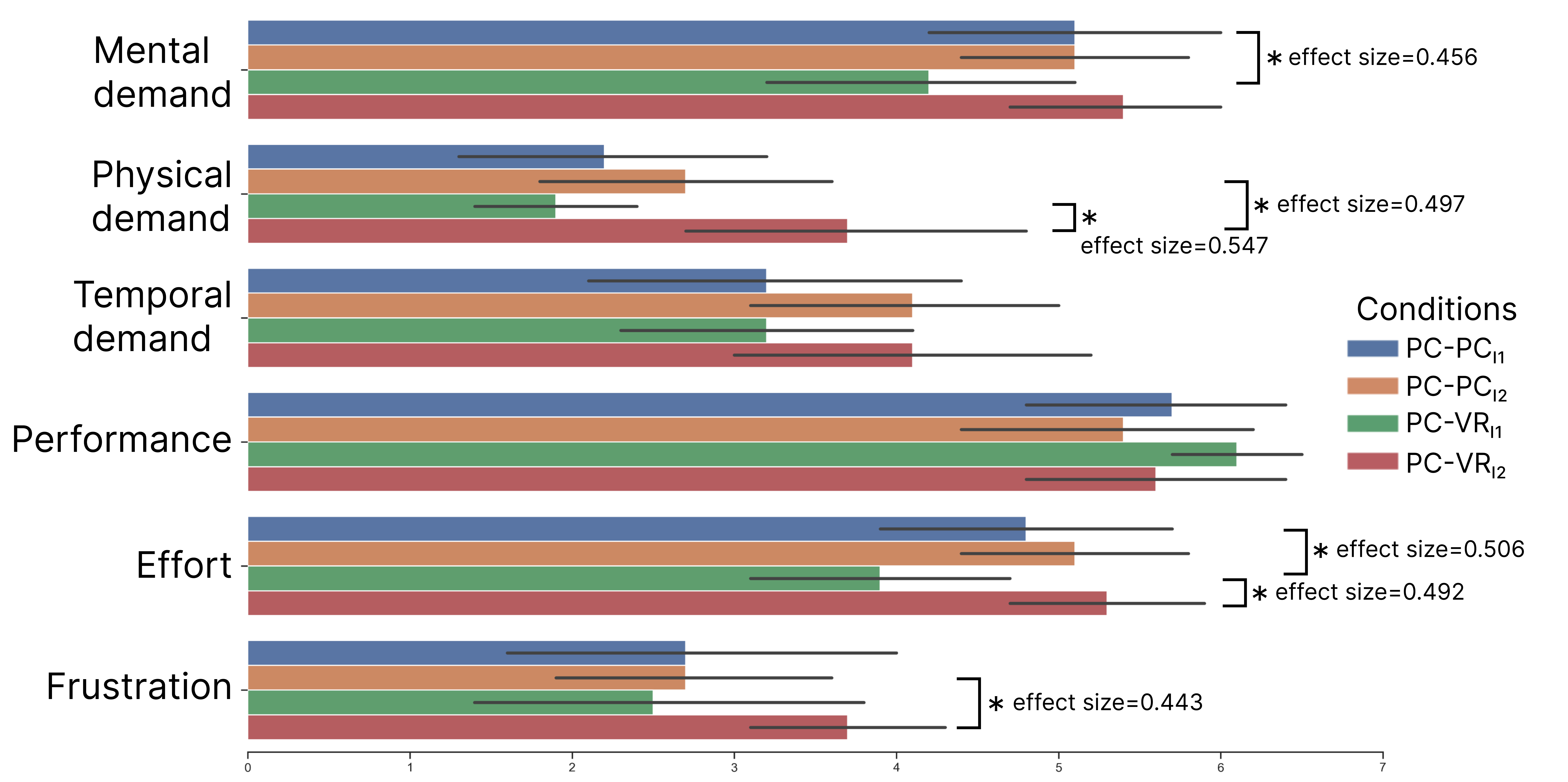}
\caption{The figure shows the mean and 95\% confidence intervals of the NASA task load questionnaire for participants in two conditions. \re{I1 represents participants who used PC in both conditions; I2 represents participants who used PC in PC-PC and VR in PC-VR. We also report the effect size calculated by the formula $\frac{Z}{\sqrt{N}}$~\cite{rosenthal1994parametric}.}{}}
\label{fig:s2_result_tl}
\end{figure}

For the perceived task load, we did not find a significant difference between PC-PC and PC-VR either. However, we find significant differences between individuals between and within conditions as shown in \autoref{fig:s2_result_tl}.
\re{For individuals between conditions, 
P$_{I1}$ found themselves had less mental demand (\pValue{0.0412}) and spent less effort (\pValue{0.0235}) in the asymmetric condition (PC-VR$_{I1}$) than in the symmetric condition (PC-PC$_{I1}$). However, P$_{I2}$ found themselves more confused (\pValue{0.0473}) and had more physical demand (\pValue{0.0260}) in asymmetric conditions (PC-VR$_{I2}$) than in symmetric conditions (PC-PC$_{I2}$).
For individuals within the asymmetric condition, P$_{I2}$ (PC-VR$_{I2}$) had more physical demand (\pValue{0.0144}) and spent more effort (\pValue{0.0278}) than $P_{I1}$ (PC-VR$_{I1}$).}{}

\re{We counted the number of operations in the interaction logs \re{(we only considered 8/10 groups data because we missed G2 and G3 interaction logs)}{}. Except for a significant increase in the average number of document retrievals in VR (\stats{137}{111}) than PC (\stats{44.4}{35.6}) (\ie, participants clicking the document on PC and participants looking at the document in VR) (\pValue{0.00854}, effect size = 0.467), we did not find significant differences in operation usage between conditions and individuals.}{}

\subsection{Communication}
We did not find a significant difference between the average number of different types of instances and behavioral engagement between symmetric and asymmetric conditions.

\subsection{\re{Preference}{UPDATE: Reorganized the format similar to Section 3}}
\re{11 participants preferred PC-VR, while nine participants preferred PC-PC.
We analyzed the reasons from their qualitative feedback, similar to the first study.}{}

\para{Symmetric collaboration.}{} The major reasons for preferring symmetric settings are allowing an equal collaboration environment and the familiarity of the environment. Four \re{participants (P3, P8, P13, and P16)}{groups} suggested that they have a smoother co-work in symmetric settings. For example, P16 appreciated that \userquote{the work was distributed equally and able to communicate the relevance of documents.} P8 complained that \userquote{I had a harder time figuring out what the other person was doing in VR.} Furthermore, \re{four participants (P3, P5, P7, P15)}{} noticed that their partners performed better in symmetric collaboration. For example, P3 commented that \userquote{my partner provided faster response when he also used a PC. For VR, he read slower than before}, and P7 stated that \userquote{the person in the VR condition might be more invested in using the tool rather than the act of collaborating in itself.}

\para{Asymmetric collaboration.}{} The major reasons for preferring PC-VR over PC-PC are allowing division of work and enhancing document reading. \re{Eight participants}{Four groups} (P1, P9, P11-12, P15, P17-18, P19) commented that asymmetric settings implicitly divided their roles and led to benefits in collaboration. For example, in G9, P17 commented that \userquote{I think the collaboration experience for the PC-VR condition was more efficient because one of the participants had a big picture for the story, and the other could focus on each document.} P18 added \userquote{In terms of collaboration, having both VR and PC, our strengths complement each other. I could see all the documents together in VR, while my partner can manage the nodes easily on PC.} P15 also stated that \userquote{since I understand VR users have some limitations (cannot type custom labels), so I would like to help him more, leading to more collaboration. ... I think it is good because I am more motivated to collaborate.} 

Moreover, three participants (P6, P10, P18) who used VR appreciated the spatial usage for viewing the documents and \re{graph visualization}{graph}. For example, P18 mentioned that \userquote{VR environment allowed me to see all the documents together. It also allowed me to easily remember which document is of interest and where it is located.} P6 further supplemented that \userquote{Reading the different documents was easy in the VR as they were all on one screen and I was just hoping from one to another.} P10 also commented that \userquote{I enjoy working in VR more than PC because I am able to see all articles and the graph. The story is right in front of me.}
\section{Discussion}
\para{\re{Optimizing both ends improves user preference for asymmetric collaborative visualization}{Effect of different levels of asymmetry}.}
\re{From the studies, we saw optimizations for PC and VR helped increase user preference towards asymmetric collaborative visualization. Compared to the \textit{mirrored VR}
we utilized the strength of both ends by enhancing VR with spatial sensemaking, embodied interaction, and enhancing PC with typing function.
One obvious concern in our current asymmetric implementation is the added visual and interaction differences between different devices may lead to more communication effort.
However, we found that the preference for asymmetric collaborative visualization has increased from 29.2\% (7/24) to 55\% (11/20). One major reason is that VR users could find the benefits of using VR in problem-solving tasks. More importantly, they were able to apply the advantages to the task. Both participants in the group would perceive improvement during collaboration. Additionally, we believe our newly implemented awareness cues improved the implicit communication experience to supplement the potentially added communication cost.}{In the formative study, we simply used the mirrored design, a low-level asymmetric condition with almost the same interface and interaction as the PC. In the second study, we took a relatively high-level asymmetric condition, where PC and VR have more differences. Specifically, we utilized the strength of both ends by enhancing VR with spatial sensemaking and embodied interaction and enhancing PC with typing function.}

\begin{figure}
\centering
\includegraphics[width=\columnwidth]{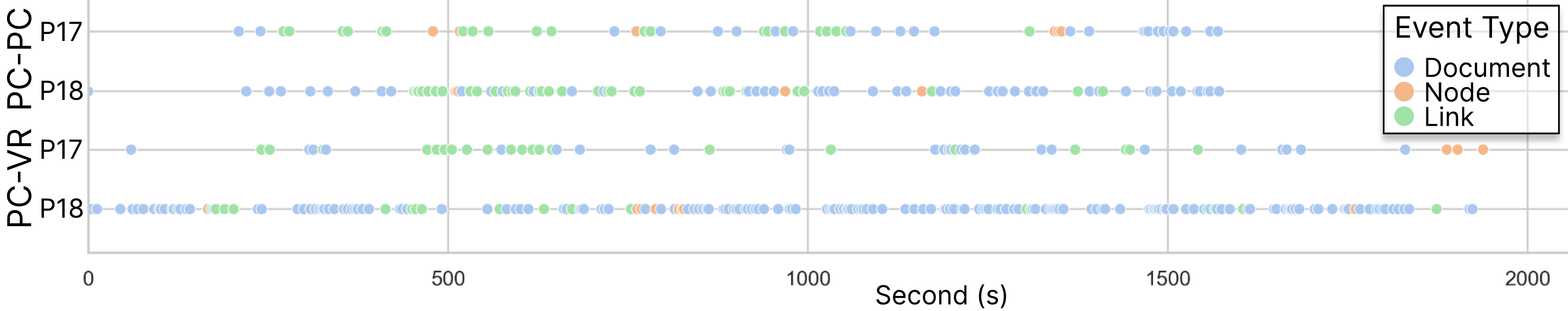}
\caption{\re{The interaction logs of G9 (P17-18). P17 focused on link operations, while P18 read documents and worked on node operations. Other groups' logs are provided as supplementary materials.}{}}
\label{fig:role}
\end{figure}

\para{Asymmetric setting did not hinder task performance and collaboration.}
Asymmetric collaborative visualization, no matter the complexity of the task and the level of asymmetry (\ie, PC-VR with the \textit{mirrored VR} or PC-VR with the refined VR) in the interface, did not affect collaboration much in terms of task performance and communication.
Based on the result, we found that there was no significant difference between asymmetric and symmetric collaboration in task completion time and accuracy in both studies, similar to a previous finding~\cite{drey2022towards}.
\re{One possible reason is that almost the same analytical functionality was given to all conditions.}{}
For communication and social behavior, we also did not see significant differences. One possible reason is that by giving similar awareness cues, participants could communicate independently of the devices.

\para{\re{Trade-offs in asymmetric collaborative visualization.}{}}
\re{We found PC users perceived significantly less mental load and effort in completing the task in PC-VR than PC-PC from the second study.}{We did find significant benefits of asymmetric collaborative visualization compared to symmetric collaboration, though there is no significant difference in task performance and collaboration efficiency.
PC users perceived less mental load and effort in completing the task.} One possible reason is that VR users could view all documents and graphs at the same time, allowing the VR users to have a better sense of the story and lead the discussion. It then reduced the effort for PC users to find related documents or memorize them compared to collaborating with another PC user.

\re{Although we can leverage the benefits from different devices in asymmetric collaboration, we need to tackle challenges from both devices to provide a better experience in collaboration. For example, the physical demand in VR is always heavier than PC environment, as interactions are usually in 3D. Participants felt frustration and physical demands when using VR, which might balance off the benefit brought to the PC users.
The collaboration experience will be further limited for the PC user when the VR user is fatigued after long use.
Moreover, asymmetric collaboration, especially with more diversified interfaces and interactions, might easily cause unbalanced teamwork. As shown in the second study, the effort between collaborators in the asymmetric setting was significantly different, while that in the symmetric setting was more balanced. The uneven workload would potentially lead to a negative effect on teamwork quality~\cite{hoegl2001teamwork}.}{}

\para{User preference towards symmetric and asymmetric collaborative visualization.}
\re{User preference for symmetric or asymmetric was equally split between participants}{Symmetric and asymmetric collaborative visualization are almost equally preferred by participants}. The reason for the preference for symmetric collaboration, especially PC-PC, is that it is the most familiar collaboration setting for all participants. Participants are able to learn and perform well easily. PC users are well-trained with the mouse input and mouse support with precise selection. On the other side, asymmetric collaboration provides two different interfaces, leveraging the best of both devices for the collaboration task. Some participants found it helpful, while some participants found performance decreased with their partners. Moreover, some participants enjoyed working at the same pace and environment, while some participants appreciated the division of labor. Echoing the finding in a prior study~\cite{bach2017hologram}, the preference for a symmetric or asymmetric collaboration could depend on the combination of the individual's abilities and characteristics.

\para{Roles in asymmetric collaborative visualization.}
\re{Based on the participants' feedback in both studies and individual interaction logs in the second study, we found that the asymmetry may motivate the division of roles in the teamwork. In particular, four participants in the first study and four participants in the second study commented that}{} the asymmetry encouraged them to divide their roles in the problem-solving process. For example, \re{under the PC-VR condition of G9 (\autoref{fig:role}), P18 was in charge of creating nodes and P17 focused on manipulating links. But P17 and P18 did not have this pattern in the PC-PC condition.
Moreover, there are other ways participants divided their roles, \eg, one worked on the overview of the story while the other worked on details. This finding echoes Chung~\etal{}'s~\cite{chung2014visporter} vision of using different devices for different roles.}{}

\para{Limitation and Future work.}
This work takes the first step toward comparing collaborative visualization in asymmetric conditions (PC-VR) and symmetric conditions (PC-PC), which presents valuable insights for the VR and visualization communities.
Although we found that asymmetric collaboration is \re{feasible}{competitive and even slightly more preferred by participants}, a longitudinal study with more participants could be ideal for more profound insights. Moreover, more studies are needed to investigate the effect beyond performance, communication, and preference, such as group awareness~\cite{saffo2023eyes}. In addition, we did not investigate VR-VR in our second study due to the variance in participants' VR experience and the duration limitation. Future studies could include the VR-VR condition.
\re{
Lastly, our work only considered distributed scenarios, \ie, people collaborating in different places.
Other collaboration scenarios, such as co-located and
mixed-presence settings (mixing co-located and distributed teams)~\cite{kim2010hugin}, should be explored in future work.
The different scenarios might require a different design for sharing information. More devices, such as AR HMDs, could be also involved in mixed-presence settings.}{}

This study inspires some future directions for asymmetric collaborative visualization. Future work can further investigate whether asymmetric roles could benefit collaborative visualization since explicit asymmetric roles could increase social presence and immersiveness~\cite{harris2019asymmetry, lee2020rolevr}.
It is also interesting to deepen our understanding of whether using the same visual representation would help better communication in asymmetric settings, where our work only presents one set of visual representations (using the 2D \re{graph visualization}{graph} in PC and 3D \re{graph visualization}{graph} in VR) for asymmetric settings.
\section{Conclusion}
\re{In the paper, we explored how the asymmetric setting affects user experience in distributed environments for collaborative visualization.}{} Two studies were conducted first to gather design requirements of asymmetric collaborative visualization and then understand the effect on the performance and communication for problem-solving compared with symmetric counterparts.
We built a cross-virtuality web-based prototype that supports users to perform collaborative problem-solving using a node-link diagram.
Using quantitative and qualitative approaches to analyze the results, we found that asymmetric settings did not harm task performance and communication while bringing a low mental demand and effort to the PC users.
We also discussed the positive impact on user preferences of optimizing the interfaces of two devices for an asymmetric setting, trade-offs and preferences for visualization of asymmetric collaboration, and possible roles during asymmetric collaboration.
We hope our work could provide insights and inspire more work on collaborative visualization with different devices.

\acknowledgments{
This research is supported in part by the Hong Kong RGC Areas of Excellence Scheme Project no. AOE/E-603/18.}

\bibliographystyle{abbrv-doi-hyperref-narrow}

\bibliography{main}
\end{document}